\documentclass[12pt]{iopart}
\usepackage{color}
\usepackage{tabularx}
\usepackage{graphicx}

\begin{document}

\title[Fokker-Planck equation for chemical Langevin equations]{From chemical Langevin equations to Fokker-Planck equation: application of Hodge decomposition and Klein-Kramers equation}
\author{W Mu}
\address{Key Laboratory of Frontiers in
Theoretical Physics and Kavli Institute for Theoretical Physics
China, Institute of Theoretical Physics, The Chinese Academy of
Sciences, P.O.Box 2735 Beijing 100190, China}
\ead{muwh@itp.ac.cn}

\author{X Li}
\address{Max Planck Institute of Molecular Biology, Systems Biology group, Am M$\ddot{u}$hlenberg 1, 14476 Potsdam-Golm, Germany}

\author{Z-c Ou-Yang}
\address{Key Laboratory of Frontiers in Theoretical Physics and
Kavli Institute for Theoretical Physics China, Institute of
Theoretical Physics, The Chinese Academy of Sciences, P.O.Box 2735
Beijing 100190, China} 
\address{Center for Advanced Study,
Tsinghua University, Beijing 100084, China}
\begin{abstract}
The stochastic systems without detailed balance are common in various chemical reaction systems, such as metabolic network systems. In studies of these systems, the concept of potential landscape is useful. However, what are the sufficient and necessary conditions of the existence of the potential function is still an open problem. Use Hodge decomposition theorem in differential form theory, we focus on the general chemical Langevin equations, which reflect complex chemical reaction systems. We analysis the conditions for the existence of potential landscape of the systems. By mapping the stochastic differential equations to a Hamiltonian mechanical system, we obtain the Fokker-Planck equation of the chemical reaction systems. The obtained Fokker-Planck equation can be used in further studies of other steady properties of complex chemical reaction systems, such as their steady state entropies. 
\end{abstract}
\submitto{\JPA}
\maketitle
The nonequilibrium behaviors of a great amount of chemical reaction networks can be described by stochastic differential equations. For example, 
the kinetics of a simple chemical chain reaction, 
\begin{equation}\label{chem}
\rightarrow \mathrm{A}\rightarrow \mathrm{B} \rightarrow,
\end{equation}
can be described by a set of stochastic differential equations, namely, chemical Langevin equations~\cite{kampen07,gillespie00,xing09},
\begin{eqnarray}\label{reactionkinetic}
\frac{dx_{1}}{dt} & = & k_{0}-k_{1}x_{1}+\frac{\sqrt{k_{0}}}{\sqrt{\Omega}}\zeta_{1}-\frac{\sqrt{k_{1}x_{1}}}{\sqrt{\Omega}}\zeta_{2},\nonumber\\
\frac{dx_{2}}{dt} & = & k_{1}x_{1}-k_{2}x_{2}+\frac{\sqrt{k_{1}x_{1}}}{\sqrt{\Omega}}\zeta_{2},\end{eqnarray}
where $x_{i}$ denotes the concentrate of $i$th species in the chemical reaction system
of interest, here, $x_1\equiv [A]$ and $x_2\equiv [B]$ ("[ ]" denotes concentrate of the specie). The $k_{i}$ are rate constants for corresponding reaction steps, $i=0,1,2$ stands for the reactions in Eq.~(\ref{chem}) from left to right. The $\Omega$ is the closed system's volume, $\zeta_i,\,i=1,2$ refers to stochastic varible.  

We use the state variables $q_{i}\equiv \ln x_{i},\,$ instead of $x_{i}\in \mathcal{R}^{+}$ to make sure $q_{i}\in \mathcal{R}$. Here, we do not require the detailed balance condition to be satisfied in our complex chemical reaction system. In general, chemical Langevin equations are nonlinear stochastic differential equations.  

It is convenient to rewrite Eq. (\ref{reactionkinetic})
to a concise form. From the point of view of tensor field theory~\cite{wald84}, Eq. (\ref{reactionkinetic}) can be written as,  
\begin{equation}\label{kinetic}
\dot{q}_{a}=w_{a}(q)+h_{ab}(q)\zeta^{b},
\end{equation}
here, $x_{a},\,\zeta_{a}$ and $w_{a}(q)$ are all $(0,1)$ tensor fields, or $1-$form in $\mathcal{R}^n$, in the language of differential form theory. The $h_{ab}$ is a symmetric $(0,2)$ tensor field, $h_{ab}=h_{ba}$. Here, we have used Penrose's abstract index notation~\cite{wald84}, for example, $\zeta_{a}=\zeta_{\mu}\left(e^{\mu}\right)_{a}$, $\zeta_{\mu},\,\mu=1,2,\ldots,n$ are the components of the $\zeta_a$
in the basis $\left\{\left(e^{\mu}\right)_a\right\}$. In particular, if use the special basis 
$\left\{\left(e^{\mu}\right)_a=\left(\mathrm{d}x^{\mu}\right)_a\right\}$, which are the dual basis of natural basis $\{\left(\partial_{\mu}\right)^a\}$, $\zeta_{a}=\zeta_{\mu}\left(dx^{\mu}\right)_{a}$. The following discussions can be applied to the systems with a large number of species, or mathematically, the systems have high-dimensional state spaces (manifolds) $\mathcal{R}^n$, with coordinates $\{q^{\mu}\},\,\mu=1,2,\ldots n$. Note that Eqs.~(\ref{reactionkinetic}) have broad applications in the researches of physical, chemical, ecological, and financial systems, Eq.~(\ref{reactionkinetic}) is not merely restricted to chemical reaction systems~\cite{cobb81,hanggi90}.   

In previous studies~\cite{ao04,yin06}, Eq.~(\ref{reactionkinetic}) was treated in the mathematical language of vectors and matrices. However, in order to effectively exploit the power results of modern mathematics, in present work, we will treat the chemical Langevin equations in a different way. In the studies on the existence of nonequilibrium processes described by stochastic different equations, Ao et al.~\cite{ao04,yin06,ao07,ao08} introduced an "axillary matrix" $\mathbf{M}(q)=\mathbf{S}(q)+\mathbf{A}(q)$, and $\mathbf{S(T)}(q)$ is symmetric (antisymmetric) part of matrix $\mathbf{M}(q)$. The meaning of the the approach can be much clearer using the language of differential form theory. By introducing an axillary tensor field $M_a^{\,\,b}$, original differential $1$-form $w_a$ in Eq.~{\ref{kinetic}) is converted to another differential $1$-form $\tilde{w}_a=M_a^{\,\,b}\,w_b$. The existence of potential means 
\begin{equation}\label{integrable}
\tilde{w}_a=(\mathrm{d}f)_a,\quad f\in C^{\infty}(U),
\end{equation}
and $U$ is open in $\mathcal{R}^n$. The necessary conditions of Eq.~(\ref{integrable}) can be expressed as following equations~\cite{wstenholtz78},
\begin{equation}\label{neccond}
\frac{\partial \tilde{w}_{\mu}}{\partial x^{\nu}}=\frac{\partial \tilde{w}_{\nu}}{\partial{x^\mu}},\quad(\mu,\,\nu=1,2,\ldots,n). 
\end{equation}

The Eq.~(\ref{neccond}) restricts the choices of axillary tensor field $M_a^{\,b}$. It is worth noting that Eq.~(\ref{neccond}) can not ensure that the potential function is $f\in C^{\infty}(U)$.   

In present work, we directly decompose the differential $1$-form $w_a$ using Hodge decomposition theorem, instead of introducing an axillary tensor field. 
The Hodge decomposition theorem is the extension of Helmholtz decomposition theorem in traditional vector analysis of $\mathcal{R}^3$. It tells us, for any differential $p$-form $w_{p},\,(p<n)$ in $\mathcal{R}^n$, it can be uniquely decomposed into three differential forms,
\begin{equation}\label{hodgetheorem}
w_{p}(q)=d\alpha_{p-1}(q)+\delta\beta_{p+1}(q)+\gamma_{p}(q),
\end{equation}
 
Here, the $\mathrm{d}$ and $\delta$ denote external derivative and codifferential operators 
respectively~\cite{wstenholtz78}. The $\alpha$ is a differential $(p-1)-$ form, $\beta$ is a differential $(p+1)-$ form, and $\gamma$ is a harmonic $p-$form,
i.e., $\Delta\gamma=(d\delta+\delta d)\gamma(q)=0$, and "$\Delta$" is the Laplace-Beltrami operator.

In particular, Hodge decomposition of differential $1$-form in $\mathcal{R}^3$ is equivalent to Helmholtz decomposition in traditional vector analysis: given a vector $\vec{V}$, it can be uniquely decomposed as $\vec{V}=\nabla\cdot\phi+\nabla\times\vec{A}$, therefore Hodge decomposition can also refer to Hodge-Helmholtz decomposition.

For our problem of interest, $1$-form $w_{a}$ can be decomposed as $w_{a}=(d\phi)_{a}+(\delta\beta)_{a}+\gamma_{a}$,
with the $\phi$ being a scalar field ($0$-form), $\beta_{ab}$ being a 2-form (antisymmetric (0,2) tensor field) and $\gamma_{a}$ being a harmonic $1$-form, i.e., $(d\delta+\,\delta\, d)\gamma_{a}=0$.

In present problem, the index of the tensor field can be lowered and raised by metric tensor in $\mathcal{R}^n$, $g_{ab}=\delta_{ab}$. Using Hodge decomposition theorem, after
certain mathematical deriving, we obtain a stochastic differential equations with a form,
\begin{equation}\label{afterhodge}
\dot{q}_{i}=-\frac{\partial\phi(q)}{\partial q^{i}}+\frac{\partial f_{ij}(q)}{\partial q^{j}}+A_{i}(q)+h_{ij}(q)\zeta^{j},\quad i=1,2,\ldots,n.
\end{equation}
Here, "$\dot{q}_i$" denotes time derivative of $q_i(t)$. The $f_{ij}$ is antisymmetric, i.e., $f_{ij}=-f_{ji}$. The constraint on $A_i$, $\partial^{2}\, A_{i}(q)=0$ implies $A_{i}(q)$ is a harmonic function. We will use the properties of harmonic function later. Here, we have already adopted Einstein summation convention. 
In Ref.~\cite{ao04}, with axillary matrix $S(\mathbf{q}_t)$ and $A(\mathbf{q}_t)$, the so called $\phi$-decomposition need further constraints, and the constrained $\phi$-decomposition is then called "gauged" $\phi-$decomposition. In our approach, the decomposition is definitely unique, which is guaranteed by Hodge decomposition theorem.  

To obtain the Fokker-Planck equation of our complex chemical reaction systems,
a powerful method is the mapping from chemical Langevin equations to a Hamiltonian mechanical system~\cite{ao04,yin06,ao07}. The deterministic forces and stochastic forces in Eq.~(\ref{afterhodge}) implies there are two different time scales in the system, which describe fast stochastic motions and slower chemical kinetic evolutions respectively. As discussed in Ref.~\cite{ao04}, slow motion can be thought of as the consequence of the inertial of the chemical reaction systems, which can be described by an inertial mass $m$ in the theory. 

We improved the approaches presented in Ref.~\cite{yin06}, map the chemical Langevin equations to a Hamiltonian system which has a $2n-$dimensional phase space, then obtain the Klein-Kramers equation for time evolution of probability distribution function. From now on, we will not use previous mathematical language of differential form, but directly use the components of differential forms, such as $p_i$, $f_{ij}$, \emph{etc}. To construct the $2n$-dimensional phase space, or so called "enlarged state space", we introduce dynamic momentum $p_i$, let 
\begin{equation}\label{H1}
{\dot{q}}_i=p_i/m, \quad i=1,2,\ldots,n.
\end{equation}

The other $n$ Hamiltonian equations are constructed based on Eq.~(\ref{afterhodge}),
\begin{equation}\label{H2}
{\dot{p}}_i=-\frac{p_i}{m}-\frac{\partial\phi(q)}{\partial q^{i}}+\frac{\partial f_{ij}(q)}{\partial q^{j}}+A_{i}(q)+h_{ij}(q)\zeta^{j}.
\end{equation}

There is no It$\hat{o}$-Stratonovich dilemma in the connection between the stochastic differential equations, Eq.~(\ref{reactionkinetic}) and the Hamiltonian equations, Eq.~(\ref{H1}) and Eq.~(\ref{H2}). 

We will truncate the results at the order of $O\left(1/m\right)$, finally integrate out the freedom of dynamic momentum $p$, and obtain the   
time evolution of $\rho(q,t)$. 
For simplification, we assume a noise in Eq. (\ref{H2}) is a standard Gaussian white noise with
$m$ independent components $\zeta_i,\,i=1,2,\ldots,m$. Define $\xi_i(q)=h_{ij}\,\zeta^j,\,i=1,2,\ldots,n,\,j=1,2,\ldots,m$, we have $\langle \xi_{i}\rangle=0$, 
and
\begin{equation}
\langle\xi_{i}(t)\xi_{j}(\tau)\rangle\,=\,2D(q)\delta_{ij}\delta(t-\tau),\quad i,j=1,2,\ldots,n.
\end{equation}
Here, the number of stochastic variables $m$ can be different than that of state variables, the matrix $D(q)$ is diffusion matrix.   

The generalized Fokker-Planck equation, so called Klein-Kramers equation can be obtained from Eq.~(\ref{H1}) and Eq.~(\ref{H2}),
\begin{eqnarray}
\partial_{t}\rho_{1}(q,p,t) & = & \partial_{p_{i}}\left[\frac{p_{i}}{m}+\partial_{q_{i}}\phi\left(q\right)-\partial_{q_{k}}\,f_{ik}-A_{i}(q)+D_{ij}(q)\partial_{p_{j}}\right]\rho_{1}(q,p,t)\nonumber \\
 &  & -\partial_{q_{i}}\cdot\frac{p_{i}}{m}\rho_{1}(q,p,t),
\end{eqnarray}
which can be rewritten as, 
\begin{equation}\label{KK}
\left\{ \partial_{t}+\frac{p_{i}}{m}\partial_{q_{i}}+G_{i}(q)\,\partial_{p_{i}}-\partial_{p_{i}}\left[D_{ij}(q)\partial_{p_{j}}+\frac{p_{i}}{m}\right]\right\} \rho_{1}(q,p,t)=0.\label{eq1}\end{equation}
Here, $G_{i}(q)\equiv \partial_{q_{k}}f_{ik}(q)+A_{i}(q)-\phi_{q_{i}}(q)$, and $\phi_{q_{i}}\equiv\partial_{q_{i}}\phi\left(q\right)$.

To get Fokker-Planck equation, we regroup the $p-$ and $q-$ derivatives in Eq.~(\ref{KK}) into two operators $\hat{L}_1$ and $\hat{L}_2$,
\begin{eqnarray}
\hat{L}_{1} & = & \partial_{p_{i}}\left(D_{ij}\partial_{p_{j}}+\frac{p_{i}}{m}\right),\nonumber \\
\hat{L}_{2} & = & -\frac{p_{i}}{m}\partial_{q_{i}}+\left[\phi_{q_{i}}(q)-A_{i}(q)-\partial_{q_{k}}f_{ik}\right]\partial_{p_{i}},\end{eqnarray} 
and Eq.~(\ref{eq1}) becomes, 
\begin{equation}
\partial_{t}\rho(q,p,t)=\left(\hat{L}_{1}\,+\,\hat{L}_{2}\right)\rho_{1}(q,p,t).
\end{equation}

Then use Gardiner's standard projection operator method to eliminate
the fast degrees of freedom of $q$ implied in the zero mass limit, or "over-dumping" limit. We introduce a projection operator $\hat{P}$, as
\begin{equation}\label{projectionoperator}
\hat{P}h(q,p,t)=\frac{1}{\left(\sqrt{2\pi m\det D}\right)^{n}}\,\exp\left(-\frac{p_{i}D_{ij}^{-1}p_{j}}{2m}\right)\int\, h(p',q',t)\mathrm{d}^{n}p'.
\end{equation}
here $h(q,p,t)$ is an arbitrary function, and $\mathrm{det}D$ is the determinant of the diffusion matrix $D(q)$.

Projection operator is an idempotent operator, $\hat{P}^2=\hat{P}$.
Notice the fact, 
\begin{equation}\label{fact}
\left(D_{ij}\partial_{p_{j}}+\frac{p_{i}}{m}\right)\exp\left(-\frac{p_{k}D_{kl}^{-1}p_{l}}{2m}\right)=0,
\end{equation}
obviously there is an identity, $\hat{L}_{1}\hat{P}=0$. 
For an arbitrary $p$-independent $c_{i}(q)$, the following identity
holds, 
\begin{equation}\label{equation}
\hat{L}_{1}p_{i}c_{i}(q)\exp\left(-\frac{p_{k}D_{kl}^{-1}p_{l}}{2m}\right)=-\frac{1}{m}p_{i}c_{i}(q)\exp\left(-\frac{p_{k}D_{kl}^{-1}p_{l}}{2m}\right).
\end{equation}

Thus, $\psi\equiv p_{i}c_{i}(q)\exp\left(-\frac{p_{k}D_{kl}^{-1}p_{l}}{2m}\right)$
is the eigenfunction of operator $\hat{L}_{1}$, with the eigenvalue
being $-1/m$, and the inverse operator $\hat{L}_{1}^{-1}$
satisfies, 
\begin{equation}
\hat{L}_{1}^{-1}\psi=-m\psi.
\end{equation}

Apply projection operator $\hat{P}$ to the phase space probability distribution function $\rho(q,p,t)$ 
\begin{equation}
\hat{P}\rho_{1}(q,p,t)=\frac{1}{\left(\sqrt{2\pi m\mathrm{det}D}\right)^{n}}\exp\left(-\frac{p_{i}\, D_{ij}^{-1}\, p_{j}}{2m}\right)\,\rho(q,t)\,\equiv v(q,p,t),
\end{equation}
with $\rho(q,t)\equiv\int\mathrm{d}p\,\rho_{1}(q,p,t)$ being the state space probability distribution. 
 
It has been proved~\cite{yin06} 
\begin{equation}\label{fenjie}
\partial_{t}\, v(q,p,t)=-PL_{2}\hat{L}_{1}^{-1}\hat{L}_{2}v\,+\, O\left(\sqrt{m}\right).
\end{equation}

Substitute of explicit forms of operators $\hat{L}_1$ and $\hat{L}_2$ to Eq.~(\ref{fenjie}), we have 
\begin{equation}
-\hat{P}L_{2}\hat{L}_{1}^{-1}\hat{L}_{2}v = \partial_{q_{i}}\left[D_{il}\partial_{q_{l}}+A_{i}(q)+\partial_{q_{i}}\phi(q)\right]v(q,p,t).
\end{equation}

Finally, by integrating out the degrees of freedom of $p$ in $\rho(q,p,t)$, we get the multivariate Fokker-Planck equation with the form,
\begin{equation}\label{FP}
\partial_{t}\rho(q,t)=\partial_{q_{i}}\left[D_{ij}\partial_{q_{j}}+A_{i}(q)+\partial_{q_{i}}\phi(q)\right]\rho(q,t).
\end{equation}

We have known $A_i(q)$ is a harmonic function. Assuming it is bounded at the state space $\mathcal{R}^n$, it must be constant, (denoted as $A_i(q)=a_i$), according to the Liouville's theorem in harmonic function theory~\cite{axler01}. With this assumption, our Fokker-Planck equation can be further reduced to 
\begin{equation}
\partial_{t}\rho(q,t)=\partial_{q_{i}}\left[D_{ij}(q)\partial_{q_{j}}+\partial_{q_i}\bar{\phi}\right]\rho(q,t).
\end{equation}
Here, $\bar{\phi}(q,t)\equiv \phi(q,t)+a_i\,q^i$, is the potential landscape of the chemical reaction systems, therefore the ''force'' has a form of 
\begin{equation}
\bar{f}_{i}=-\left[A_{i}(q)+\partial_{q_{i}}\phi(q)\right]=-\left[a_{i}+\partial_{q_{i}}\phi(q)\right]=-\partial_{q_{i}}\bar{\phi}(q).
\end{equation}

To conclude, we have used Hodge decomposition approach to study the potential landscape for chemical reaction systems described by chemical Langevin equations, which are stochastic differential equations. We do not require detailed balance condition to be held in the systems. Hodge decomposition theorem is very convenient to deal with the problem related to high-dimensional state space. We map the chemical Langevin equations to a Hamiltonian mechanical system, and find the Fokker-Planck equation for chemical Langevin equations. With certain assumptions, there does exist potential landscape, the potential can also be time-dependent. Our work can be used in the studies of other steady state properties of the chemical reaction systems, such as their steady state entropies.  



\section*{References}
\begin{thebibliography}{999}
\bibitem{kampen07}
van Kampen N G 2007 \emph{Processes in Physics and Chemistry} (3rd Ed. North Holland).

\bibitem{gillespie00}
Gillespie D T 2000, \emph{J. Chem, Phys.} {\bf 113}, 297.

\bibitem{xing09}
Xing J 2009, eprint arXiv:0908.4526.

\bibitem{wald84} There are many modern textbooks on general relativity
adopting Penrose's abstract index notation, e.g., Wald R M 1984 \emph{General
Relativity}, pp. 23, (University of Chicago Press, Chicago
and London). 

\bibitem{cobb81}
Cobb L, and Thrall R M D 1981 \emph{Mathematical Frontiers of the Social and Policy Sciences} (Westview Press, Boulder, Colorado).

\bibitem{hanggi90}
Hanggi P, Talkner P, and Bokovec M 1990 \emph{Rev. Mod. Phys.} {\bf 62}, 251.

\bibitem{wstenholtz78}
Von Westenholtz C 1978, \emph{Differential
forms in mathematical physics}, (North-Holland Publishing Company, Amsterdam).

\bibitem{ao04}
Ao P 2004 \emph{J. Phys. A: Math Gen.} {\bf 37}, L25-L30.

\bibitem{yin06}
Yin L, Ao P 2006 \emph{J. Phys. A: Math. Gen.} {\bf 39}, 8593-8601. 

\bibitem{ao07}
Ao P, Kwon C, and Qian H 2007 \emph{Complexity}, {\bf 12}: 19-27. 

\bibitem{ao08}
Ao P 2008 \emph{Commun. Theor. Phys.} {\bf 49}, 1073-1090.

\bibitem{gardiner04}
Gardiner C W 2004 \emph{Handbook of Stochastic Methods} (3rd Ed. Springer, Berlin)

\bibitem{zwanzig01}
Zwanzig R 2001 \emph{Nonequilibrium Statistical Mechanics} (Oxfored University Press).

\bibitem{kubo92}
Kubo R, Toda M and Hashitsume N 1992 \emph{Statistical Physics II} (2nd Ed. Springer, Berlin).

\bibitem{axler01}
Axler S, Bourdon P, and Ramey W 2001 \emph{Harmonic Function Theory}, (Springer-Verlag New York, Inc.)


\end{thebibliography}
\end{document}